\documentclass[conference]{IEEEtran}
\IEEEoverridecommandlockouts
\usepackage{cite}
\usepackage{listings}
\usepackage{amsmath,amssymb,amsfonts}
\usepackage{algorithmic}
\usepackage{graphicx}
\usepackage{textcomp}
\usepackage{xcolor}
\usepackage{makecell}
\usepackage{caption}
\usepackage{float}
\usepackage{tikz}
\newfloat{lstfloat}{htbp}{lop}
\floatname{lstfloat}{Listing}
\usepackage[T1]{fontenc}

\lstdefinelanguage{mlir}{
    alsodigit = {.},
    keywords = {stencil.apply, stencil.access, arith.constant, arith.addf, arith.mulf, stencil.return, f32, f64, stencil.index,stencil.temp, hlfir.declare, fir.load, arith.cmpi, fir.if, fir.result, scf.if, scf.yield, func.return, fir.alloca, hlfir.assign, memref.alloca, memref.store, memref.load, memref.alloc, arith.subi, arith.addi, linalg.yield, linalg.reduce}
}

\def\BibTeX{{\rm B\kern-.05em{\sc i\kern-.025em b}\kern-.08em
    T\kern-.1667em\lower.7ex\hbox{E}\kern-.125emX}}
\begin{document}

\title{Fully integrating the Flang Fortran compiler with standard MLIR}

\author{\IEEEauthorblockN{Nick Brown}
\IEEEauthorblockA{\textit{EPCC} \\
\textit{University of Edinburgh}\\
Edinburgh, UK \\
n.brown@epcc.ed.ac.uk}
}

\newcommand\copyrighttext{%
  \footnotesize For the purpose of open access, the author has applied a Creative Commons Attribution (CC BY) license to any Author Accepted Manuscript version arising}
\newcommand\copyrightnotice{%
\begin{tikzpicture}[remember picture,overlay]
\node[anchor=south,yshift=10pt] at (current page.south) {\fbox{\parbox{\dimexpr\textwidth-\fboxsep-\fboxrule\relax}{\copyrighttext}}};
\end{tikzpicture}%
}

\maketitle
\copyrightnotice

\begin{abstract}
Fortran is the lingua franca of HPC code development and as such it is crucial that we as a community have open source Fortran compilers capable of generating high performance executables. Flang is LLVM's Fortran compiler and leverages MLIR which is a reusable compiler infrastructure which, as part of LLVM, has become popular in recent years. 
However, whilst Flang leverages MLIR it does not fully integrate with it and instead provides bespoke translation and optimisation passes to target LLVM-IR. In this paper we first explore the performance of Flang against other compilers popular in HPC for a range of benchmarks before describing a mapping between Fortran and standard MLIR, exploring the performance of this. The result of this work is an up to three times speed up compared with Flang's existing approach across the benchmarks and experiments run, demonstrating that the Flang community should seriously consider leveraging standard MLIR.
\end{abstract}


\section{Introduction}
With almost 70 years of lineage in scientific computing, it is unsurprising that Fortran is ubiquitous in HPC. Indeed, around 80\% of applications that are run on ARCHER2, the UK national supercomputer, are written in Fortran \cite{rodriguez2023fortran}. Consequently, it is important that the HPC community has access to advanced Fortran compilers which can produce binaries with optimal performance making most effective use of these machines. Whilst there are many Fortran compilers, it is important that we as a community invest in open source solutions so that we can influence the technology and avoid vendor lock in. 

Flang \cite{flang} is one such open source Fortran compiler. Part of the LLVM project, the current version of Flang is a complete rewrite of the previous LLVM Fortran compiler in recent years, and aims to provide a compiler which is capable of handling the full range of the Fortran language specification, although this is yet to be realised. Flang is built upon MLIR \cite{lattner2021mlir} which is also part of LLVM and provides composable Intermediate Representations (IRs) which promotes increased sharing of compiler infrastructure. In MLIR there are a large number of IR dialects, along with transformations between dialects and optimisations. MLIR also provides a framework for developing bespoke dialects and transformations, and indeed a rich ecosystem has been developed by the community. The design of Flang is somewhat surprising in that whilst it defines its own MLIR dialects, it sits apart from the rest of standard MLIR with the exception of leveraging a number of standard dialects. Consequently, that Flang implements its own optimisations and lowerings, sitting apart from the work being developed by the community in MLIR. Not only does this increase duplication, but furthermore has the potential to impact performance because Flang is unable to take advantage of the progress being made in MLIR, much of which is driven by hardware vendors.

In this paper we explore an alternative approach, where Flang's MLIR dialects are lowered to the standard MLIR dialects, then relying on existing MLIR transformations and optimisations to build binaries. We have developed a research prototype that enables us to test the hypothesis that such an approach can deliver improved performance and help close the performance gap between Flang and other, more mature, Fortran compilers. 

This paper is structured as follows; after exploring the background to this work in Section \ref{sec:bg}, we then describe the setup used for experiments throughout this paper in Section \ref{sec:experimental}, before undertaking a performance comparison of binaries produced by Flang against the Cray and Gfortran compilers in Section \ref{sec:flang-perf-comparison}. Section \ref{sec:our-approach} describes our mapping between Fortran concepts in Flang's dialects and the standard MLIR dialects, before we explore the performance that this delivers in Section \ref{sec:performance}. Lastly, conclusions are drawn in Section \ref{sec:conclusions} which also discusses further work.

The contributions of this paper are as follows:
\begin{itemize}
  \item A performance comparison between Flang and other compilers common in HPC, helping to understand the difference in performance that these deliver.
  \item Description of a mapping between Fortran concepts and standard MLIR dialects, exploring how the major constructs in Fortran can be represented in standard MLIR. 
  \item An exploration of the performance delivered by the existing Flang approach compared to leveraging a standard MLIR flow.
\end{itemize}


\section{Background and related work}
\label{sec:bg}
LLVM \cite{lattner2004llvm} comprises set of reusable compiler and tool chain technologies for developing compilers across many different languages, targeting a rich set of hardware. LLVM provides numerous language frontends and architecture specific backends which are connected via LLVM-IR. An LLVM frontend, such as Clang for C and C++, generating LLVM-IR is therefore able to target any backend, and backends for a wide range of architectures including CPUs, GPUs, and FPGAs have been developed. However, LLVM-IR itself is low level, and this requires significant work by each of the frontends to target LLVM-IR and results in duplication between them.

MLIR, which was first developed by Google and released open source thereafter, instead provides a series of IR dialects and transformations between these. Instead of targeting the low level LLVM-IR, frontends can translate to higher level intermediate representations and then rely on common transformations within MLIR to do the rest of the work in reaching LLVM-IR. IR follows Static Single Assignment (SSA) form, and one of the major strengths of MLIR is that dialects can be mixed in the IR and manipulated separately, enabling progressive lowering of the abstraction level ultimately to LLVM-IR.  Because such lowerings involve existing dialects and transformations, this approach enables much greater sharing of compiler infrastructure between frontends, significantly reducing the overall software effort in developing compilers. MLIR also provides a framework for compiler developers to create their own dialects and transformations. 

MLIR has become popular since it became a sub-project of LLVM, and an extensive community contributes to the development of the technology. There are many IR dialects provided as standard by MLIR, including \emph{arith} for arithmetic operations, \emph{scf} for structured control flow providing serial and parallel loops, \emph{affine} which represents affine loops, \emph{memref} for memory management and data access, \emph{func} to represent functions and calling between them, \emph{openmp} for OpenMP parallelism, \emph{linalg} which express linear algebra operations, and \emph{vector} for vectorisation. All of these ultimately lower to the \emph{llvm} MLIR dialect, from which LLVM-IR can be generated by the \emph{mlir-translate} tool.

\subsection{xDSL}

One of the challenges faced by MLIR is the steep learning curve, where developer must leverage C++, understand LLVM concepts, work with the Tablegen format to describe dialects and keep track of the fast evolving MLIR repository. All this adds to significantly to the overhead involved in development. 

By contrast, xDSL \cite{xdsl} is a Python based compiler design toolkit which is 1-1 compatible with MLIR. Providing both the majority of standard MLIR dialects, as well as numerous additional experimental ones too, these are all expressed in the IRDL \cite{fehr2022irdl} format within Python classes. One of the purposes of xDSL is to enable rapid exploration and prototyping of MLIR concepts, with a view to these then being committing into the main MLIR codebase once these become mature and are proven. Indeed, the recent MPI dialect is one such example of this process, where it was first developed in xDSL \cite{bisbas2024shared} before being standardised in MLIR. As xDSL is 1-1 compatible with MLIR, one is able to arbitrarily go between the two technologies during compilation. We used xDSL to develop the transformations and optimisations described in this paper.

\subsection{Flang}

Flang \cite{flang} is LLVM's Fortran frontend which replaces the previous Flang Fortran compiler, known as classic Flang, with a ground-up rewrite built on-top of MLIR. An official component of LLVM, the objective of Flang is to support the full range of standard Fortran, including being able to adapt to future versions of the language. However at the time of writing whilst Flang is progressing rapidly, support for Fortran at or beyond 2003 is still work in progress and yet to reach full maturity.

\begin{figure}[htb]
\centering
 \includegraphics[width=\columnwidth]{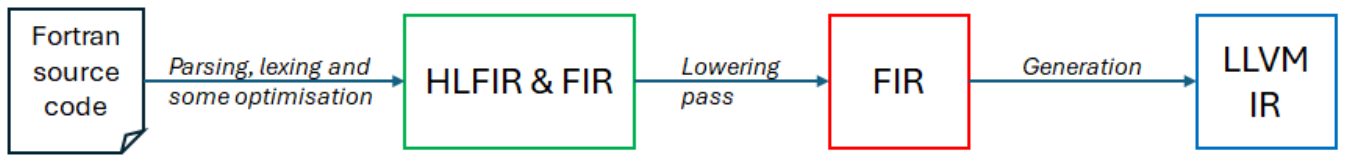}
\caption{Illustration of flow used by Flang to generate LLVM-IR from Fortran source code}	
\label{fig:flang-flow}
\end{figure}

Figure \ref{fig:flang-flow} illustrates a sketch of the overarching Flang compilation flow where, after lexing and parsing of a user's Fortran code, some optimisations are undertaken on the AST which is then lowered to Flang's High Level Fortran IR (HLFIR) \cite{hlfir} and Fortran IR (FIR) \cite{fir} dialects. These two MLIR dialects have been developed as part of Flang to specifically represent Fortran constructs and concepts. HIFIR is the newer of the two, and was developed based upon lessons learnt with FIR to provide increased abstraction over and above FIR, for instance better representing variable allocation and Fortran specific intrinsics such as linear algebra operations. These two dialects are initially mixed together, with the HLFIR components of the IR then lowered to FIR only by Flang. For example, such a lowering will leveraging the information about array bounds encoded in HLFIR and implicit in use by the HLFIR and FIR IR, to then explicitly calculate array access offsets in the lowered FIR only IR. Flang then directly generates LLVM-IR from the FIR only based IR.

In addition to HLFIR and FIR, Flang also uses a handful of standard MLIR dialects such as  \emph{arith} for arithmetic support, \emph{func} to represent functions and \emph{openmp} for threaded parallelism. However, a surprising design decision was that apart from the use of these relatively small number of standard dialects, Flang sits apart from the rest of the MLIR ecosystem. As per Figure \ref{fig:flang-flow}, LLVM-IR is generated directly from FIR, and the handful of standard MLIR dialects in use, by Flang rather than going via the rich set of existing transformations and optimisations present in MLIR. Furthermore, the HLFIR and FIR dialects are not registered with standard MLIR, so through Flang it is not possible to freely mix the Flang dialects with the standard MLIR dialects and leverage the ecosystem.

There are arguments in favour of the approach adopted by Flang, for instance decoupling from the fast moving MLIR and to be entirely confident of a flow that can encode the rich set of Fortran constructs. Indeed given that current Flang stated to be rewritten at at time when MLIR was less mature and evolving extremely rapidly, this was probably a sensible decision. However, based on the increased maturity of these technologies, better integration with standard MLIR would enable the compiler to leveraging existing MLIR passes which will undertake much of the lowering to LLVM-IR rather than having to develop these in Flang. Furthermore, such an approach would also enable Flang to benefit from the optimisation passes and tuning that is part of MLIR. 

The optimisation of Flang for stencil workloads was explored in \cite{brown2023fortran} where the authors developed a transformation which identified and extracted stencil based computations from the FIR generated by Flang. These were then represented in a stencil dialect \cite{gysi2021domain} which was lowered to LLVM-IR. This work demonstrated that whilst there was a large performance gap between the Cray and Flang compilers for such workloads, specialising using domain specific knowledge by leveraging a stencil dialect resulted in noteworthy speedups for Flang.

\section{Experimental setup}
\label{sec:experimental}
Results reported from the experiments run throughout this paper are averaged over five runs. We build and execute CPU codes on ARCHER2, a Cray-EX which is the UK national supercomputer. Each node in ARCHER2 comprises dual AMD EPYC 7742 64-core processors running at 2.25GHz, 256 GiB of DRAM and is connected to a 14.5 PB HPE Cray ClusterStor filesystem running Lustre. The AMD EPYC 7742 implements AVX2 which provides 256-bit wide vector registers. The GPU codes are built and run on Cirrus, an SGI/HPE 8600 cluster, where each GPU node is equipped with Nvidia Tesla V100-SXM2-16GB GPUs and two 2.5 GHz, 20-core Intel Xeon Gold 6148 CPUs (Cascade Lake), along with 384GiB of DRAM. All codes are built with optimisation level three, apart from the Cray compiler in Section \ref{sec:flang-perf-comparison} where we report the faster of level three or the default level two.

\section{Flang performance comparison}
\label{sec:flang-perf-comparison}
We first undertook a performance comparison of Flang against other compilers which are commonplace in HPC to identify how well these can build optimised executables. We felt that this was an important activity to help understand whether there is a performance gap between Flang and other, more mature, Fortran compilers, and hence whether it is worthwhile exploring an alternative compilation flow for Flang based upon standard MLIR. 

Table \ref{tab:performance_compilers} a performance comparison for Flang against the Cray (version 15.0.0) and GNU Gfortran (version 11.2.0) compilers which are the latest versions available on ARCHER2 and both commonplace in HPC. Indeed, we compare performance for two versions of Flang, version 20.0.0 (based upon LLVM release 18.1.8) which at the time of writing is the latest release of Flang and LLVM, and version 17.0.0 (based upon LLVM release 16.0.3) which is from May 2023. It is instructive to compare these two version of Flang, not only to understand how performance is progressing but furthermore there are significant differences in Flang between these versions because version 17.0.0 does not use HLFIR and relies on FIR only in its compilation flow. Furthermore, version 17.0.0 was the version used in mapping to the stencil dialect in \cite{brown2023fortran}.

We have selected benchmarks from the Polyhedron Fortran Benchmarks suite \cite{ftn-benchmark} which is a set of Fortran 90 programs designed to compare the performance of executables produced by different compilers. Moreover, we have included three additional benchmarks which are representative of common HPC workloads. The first of these performs a Jacobi iteration solving LaPlace's equation for diffusion, \emph{jacobi} in Table \ref{tab:performance_compilers}, working over a single two dimensional grid of size 1024 by 1024 and 100000 iterations. The second additional benchmark is the Piacsek and
Williams (PW) advection scheme \cite{piacsek1970conservation}, \emph{pw-advection} in Table \ref{tab:performance_compilers}, from the Met Office's MONC atmospheric model \cite{brown2020highly}. This scheme computes across three fields each on a three dimensional grid of size 2048 by 1024 by 1024 and  calculates the movement of quantities through the atmosphere due to kinetic effects such as wind. The third is a tracer advection scheme from the NEMO ocean model benchmarking suite \cite{psyclonebench}, \emph{tra-adv} in Table \ref{tab:performance_compilers}, which computes over six fields on a three dimensional grid of size 1024 by 512 by 512, running over 20 iterations.

\begin{table}[htb]
    \centering
    \caption{Runtime of benchmarks for versions v20 and v17 of the Flang compiler, the Cray compiler v15 and GNU Gfortran v11.2.0}
    \label{tab:performance_compilers}
    \begin{tabular}{|c|cccc|}
    \hline
        & \multicolumn{4}{c|}{\textbf{Benchmark Runtime (s)}}\\
      \textbf{Benchmark} & \textbf{Flang v20} & \textbf{Flang v17} & \textbf{Cray} & \textbf{GNU} \\
      \hline
    ac & 11.89 & 10.82 & 8.67 & 31.43\\
    aermod & DNC & 17.80 & 11.67 & 13.16 \\
    air & 5.80 & 5.15 & 3.27 & 6.88 \\
    capacita & 37.82 & 32.79 & 36.33 & 36.71 \\
    channel & 56.84 &  55.96 & 50.26 & 54.46\\
    doduc & 16.65 & 16.41 & 12.89 & 15.61\\
    fatigue & 105.90 & 111.08 & 121.57 & 99.42 \\
    gas\_dyn & 116.90 & 99.04 & 46.29 & 68.38 \\
    induct & 126.23 & 126.36 & 38.19 & 35.15 \\
    linpk & 6.24 & 5.84 & 5.79 & 4.81 \\
    mdbx & 11.37 & 12.40 & 9.19 & 12.68 \\
    mp\_prop\_design & 120.71 & 118.10 & 30.10 & 216.00 \\
    nf & 10.29 & 14.16 & 7.72 & 7.43 \\
    protein & 33.06 & 35.79 & 30.82 & 26.82 \\
    rnflow & 27.22 & 29.32 & 15.31 & 44.00 \\
    test\_fpu & 110.80 & 267.68 & 32.56 & 76.99 \\
    tfft & 48.90 & 53.98 & 61.65 & 115.86 \\
    jacobi & 277.67 & 301.92 & 109.89 & 232.62\\
    pw-advection & 205.33 & 602.43 & 47.28 & 192.05\\
    tra-adv & 141.95 & 145.82 & 79.38 & 116.71\\
    \hline
    \end{tabular}
\end{table}

It can be seen from Table \ref{tab:performance_compilers} that, whilst there are some outliers, the Cray compiler generally produces the most efficient executables. Whilst the performance delivered by the GNU and Flang compilers is more mixed, on average the GNU compiler tends to build executables with higher performance, although it is much closer than with the Cray compiler and there are numerous exceptions. For the \emph{jacobi}, \emph{pw-advection} and \emph{tra-adv} benchmarks, which are all stencil based, the Cray compiler delivers significantly better performance than the Flang or GNU compilers, with Flang producing the lowest performing executables. It should be noted that \emph{DNC} for the \emph{aermod} benchmark when using Flang v20 is because this benchmark did not compile for that version of Flang. After approximately thirty minutes of compile time, the compiler incorrectly reported that a variable had not been defined. Comprising 51888 lines, this benchmark is an extremely long file and it is likely that this is not a common use-case envisaged by the developers or regularly tested against. Interestingly Flang v17 did compile this specific benchmark, although delivered lower performance than the other two compilers.

It is instructive to explore specific outliers where Flang performs considerably better, or worse, than the other compilers. We profiled the \emph{tfft} benchmark, where Flang outperformed other compilers, exploring properties of executables produced by the Flang and GNU compilers. The entirety of the floating point for this workload is single precision, and whilst Flang was unable to vectorise any of this, the GNU compiler undertook 128-bit vectorisation of 47\% of these floating point operations. However, the executable produced by the GNU Gfortran compiler stalled 68\% of the time due to memory bound related issues and only 22\% of its instructions were floating point operations. By contrast, the executable produced by the Flang compiler stalled 51\% of the time due to being memory bound, and 27\% of instructions were floating point. Therefore, for this benchmark the Flang compiler was able to better structure memory accesses to avoid bottlenecks which resulted in a greater percentage of floating point instructions, albeit all of which were entirely scalar. 

By contrast, for the \emph{induct} benchmark the Flang compiler performed considerable worse than the Cray or GNU compilers. From profiling, it was observed that the GNU compiler resulted in 60\% of all instructions being double precision floating point, of which 67\% were 128-bit vectorised. The Flang compiler produced an executable where 58\% of the instructions were double precision floating point, however all of these were scalar. The largest discrepancy was in the number of instructions issued between these two executables, whereas the GNU produced executable issued 383 billion instructions, the Flang produced executable issued 704 billion, which is likely in large part due to the lack of vectorisation.

From exploring these performance profiles, we observe that the Flang compiler struggles to vectorise floating point operations, by far most commonly producing executables that comprise entirely of scalar operations. However, where other factors such as memory access contribute more to the overall runtime, then this becomes less of an issue and potentially the Flang compiler can deliver comparable or better performance.

We also explored the differences in performance between version 20 and version 17 of the Flang compiler. From the results in Table \ref{tab:performance_compilers} it can be observed that this difference is heavily benchmark dependent. For instance, for some benchmarks such as \emph{ac}, \emph{capacita} and \emph{gas\_dyn} the older version of Flang produces exectuables that deliver better performance than the latest version. However, there is a marked performance improvement with version 20 for the \emph{test\_fpu}, \emph{jacobi} and \emph{pw-advection} benchmarks. The narrative around performance between these versions is therefore rather nuanced, and whilst recent changes in using HLFIR have clearly benefited some workloads, others have suffered.

\section{A mapping between Flang and standard MLIR}
\label{sec:our-approach}
\begin{figure*}[htb]
\centering
 \includegraphics[width=\textwidth]{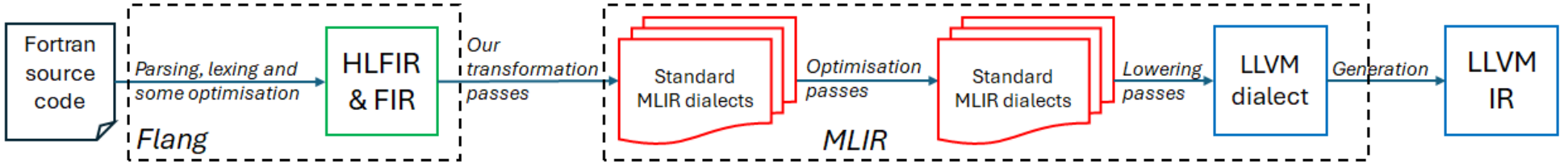}
\caption{Illustration of our flow which lowers Flang's HLFIR and FIR dialects to the standard MLIR dialects, and then uses existing transformation and optimisation passes to ultimately generate LLVM-IR. Our transformation passes are developed in xDSL, generating IR that is consumed by MLIR.}	
\label{fig:our-flow}
\end{figure*}

Figure \ref{fig:our-flow} illustrates the approach that we have adopted in this paper to map the MLIR dialects of Flang to standard MLIR. We still leverage the components of Flang which undertake lexing, parsing and initial optimisation of the code, but we then intercept the combined HLFIR and FIR based IR that is generated. We developed a transformation pass that then lowers these Flang dialects to the standard MLIR dialects, and our flow then leverages existing standard MLIR transformation and optimisation passes to lower to the LLVM MLIR dialect. LLVM-IR is then generated via the \emph{mlir-translate} tool. Our approach comprises the mapping between Flang's dialects and standard MLIR, along with several optimisation passes that are described later in this section and Section \ref{sec:performance}, ultimately providing a Flang based solution that can exploit the transformations and optimisations that are part of MLIR. Our transformation pass is developed in xDSL, which generates IR that is consumed by MLIR.

Listing \ref{lst:mlir_pipeline} illustrates the base MLIR pipeline pass that we use to lower the standard dialects into the LLVM dialect, although based upon the optimisations that we identify and discuss in Section \ref{sec:performance} this is tweaked with additional passes as appropriate. After undertaking \emph{mlir-translate} to generate the corresponding LLVM-IR, we then use clang to generate an object file and link against any other required object files and libraries including Flang's runtime libraries.

\begin{lstlisting}[language=bash, frame=lines, label=lst:mlir_pipeline, numbers=none, caption=Command issued to lower standard dialects to LLVM dialect via \emph{mlir-opt}]
mlir-opt --pass-pipeline="builtin.module(canonicalize, cse, loop-invariant-code-motion, convert-linalg-to-loops, convert-scf-to-cf, convert-cf-to-llvm{index-bitwidth=64}, fold-memref-alias-ops,lower-affine,finalize-memref-to-llvm, convert-arith-to-llvm{index-bitwidth=64}, convert-func-to-llvm, math-uplift-to-fma, convert-math-to-llvm, fold-memref-alias-ops,lower-affine,finalize-memref-to-llvm, reconcile-unrealized-casts) input_file.mlir
\end{lstlisting}

\subsection{Handling of control structures}
\label{sec:ctrl}
Many of the control structures expressed in Fortran such as conditionals, and represented by the HLFIR and FIR IR dialects by Flang, map naturally to operations in MLIR's structured control flow (scf) dialect. Listing \ref{lst:if_example_fir} is an example of Flang's representation, in HLFIR and FIR, of a Fortran subroutine which contains an if condition checking whether a variable, that itself is provided as an input to the subroutine, equals 50. It can be observed that Flang uses the \emph{func.func} operation (the \emph{func} operation in the \emph{func dialect}) to represent Fortran procedures, and at line one the function accepts a reference to a 32 bit integer (\emph{!fir.ref<i32>}) identified as \emph{\%arg0}. This argument is provided as an operand to the \emph{hlfir.declare} operation at line two which it can be seen annotates additional variable information, such as the intent of the argument in this case. It is the result of this \emph{hlfir.declare} operation that is then used throughout the rest of the IR when referencing this variable. The value held in the variable is loaded at line three, with a constant value of 50 defined at line four via the \emph{arith.constant} operation. The \emph{arith.cmpi} operation at line five performs an integer comparison between these two values, with the \emph{eq} property defining the type of comparison to undertake. This comparison returns a boolean value, represented in MLIR as type \emph{i1}, which is used as an input to the \emph{scf.if} operation at line six. A conditional operation contains two blocks, the first at lines seven and eight represent the true instructions to execute, and the block at line ten the else instructions. It can be seen that each of these blocks is terminated by the \emph{fir.result} operation which is required even if the block is empty. Lastly in Listing \ref{lst:if_example_fir}, the \emph{func.return} operation at line twelve terminates the function, in this case returning no value. This IR illustrates that whilst Flang's IR comprises heavily of operations and attributes (types) from the HLFIR and FIR dialects, it also mixes in a small number of other MLIR dialects such as \emph{arith} and \emph{func} where required.

\begin{lstlisting}[language=mlir, frame=lines, label=lst:if_example_fir, numbers=left, caption=HLFIR and FIR based IR generated by Flang for a subrouting with a conditional]
"func.func @_QMproblem_modPrun_solver(%arg0 : !fir.ref<i32> {"fir.bindc_name" = "i"}) {
    %1, %2 = hlfir.declare(%arg0) <{"fortran_attrs" = #fir.var_attrs<intent_in>}> : (!fir.ref<i32>) -> (!fir.ref<i32>, !fir.ref<i32>)
    %3 = fir.load(%1) : (!fir.ref<i32>) -> i32
    %4 = arith.constant 50 : i32
    %5 = arith.cmpi eq, %3, %4 : i32
    "fir.if"(%5) ({
        ...
        "fir.result"() : () -> ()
    }, {
        "fir.result"() : () -> ()
    }) : (i1) -> ()
    "func.return"() : () -> ()
}
\end{lstlisting}

Listing \ref{lst:if_example_mlir} illustrates the corresponding IR which has been lowered by our transformation from that of Listing \ref{lst:if_example_fir} to the standard MLIR dialects. In Listing \ref{lst:if_example_mlir} the argument to the function is represented by the identifier \emph{\%1} and this is used directly in the IR. Because in Listing \ref{lst:if_example_fir} this was a scalar with intent \emph{in}, our transformation can lower the argument directly to a scalar value rather than a reference or pointer. The same \emph{arith.constant} and \emph{arith.cmpi} operations are used as is in Flang's produced IR, with an \emph{scf.if} operation which consumes the result of the conditional. The \emph{scf.if} operation is semantically very similar to \emph{fir.if}, with \emph{scf.yield} terminating both blocks at lines six and eight. It can be seen that there are a large number of similarities between Listing \ref{lst:if_example_fir} and \ref{lst:if_example_mlir}, with the major difference being that our transformation has interrogated the information provided as part of the \emph{hlfir.declare} operation in the Flang IR, and used this to determine the most appropriate way of representing the function argument.

\begin{lstlisting}[language=mlir, frame=lines, label=lst:if_example_mlir, numbers=left, caption=Standard MLIR dialect IR generated by our lowering from Flang's HLFIR and FIR based IR]
func.func @"_QMproblem_modPrun_solver"(%1 : i32 {"bindc_name" = "i"}) {
    %2 = arith.constant 50 : i32
    %3 = arith.cmpi eq, %1, %2 : i32
    "scf.if"(%3) ({
        ... 
        scf.yield
    }, {
        scf.yield
    }) : (i1) -> ()
    func.return
}
\end{lstlisting}

Not all Fortran constructs enable such a straightforward mapping as conditionals do. For instance, Fortran do loops are represented in FIR by the \emph{fir.do\_loop} operation and our transformation maps these to the \emph{scf.for} operation. However, Fortran do loops can iterate backwards when provided with a negative step value, whereas MLIR's \emph{scf.for} operation only supports forwards iteration and the step value must be positive. Consequently, our transformation pass identifies whether the step is positive, negative or unknown at compile time. If the step is negative then loop bounds provided to \emph{scf.for} are reversed, as the loop is counting downwards in \emph{fir.do\_loop} whereas in \emph{scf.for} it must count upwards, and the absolute value of the step provided to \emph{scf.for}. If the step is unknown at compile time, then a runtime check using \emph{scf.if} is included in the code. The way in which Flang represents a do loop in \emph{fir.do\_loop} is that the first operation in the loop body is an assignment of the loop index to the loop iteration variable and it is this variable that is used to refer to the loop index by operations throughout the rest of the loop. Therefore, if the do loop is counting downwards in the generated IR we use the upwards counting index of the \emph{scf.for} loop to calculate the corresponding downwards index and store this.

Do loops in FIR are similar to {scf.for} loops in that the only way in which they can exit is based upon the loop count. However, Fortran keywords such as exit or goto can leave the loop early. Therefore, FIR also contains a \emph{fir.iterate\_while} operation, which both counts from an initial to final value but also checks a boolean flag each iteration to determine whether to continue iterating or not. This maps to the \emph{scf.while} operation, but more work required as \emph{scf.while} iterates only until a boolean condition is false as evaluated by the \emph{scf.condition} operation. Our lowering therefore adds an explicit loop variable, which is incremented each iteration and the \emph{arith.and} operation to take the logical \emph{and} of the boolean exit condition and whether the loop index has reached its terminating value. 

In addition to structured control flow, Flang's IR also jumps directly between blocks for instance to represent \emph{goto}. In doing this Flang leverages the control flow (cf) dialect, and specifically \emph{cf.br} to branch and \emph{cf.cond\_br} for a conditional branch. The \emph{cf} dialect is one of the standard MLIR dialects, so this is a straightforward conversion apart from the fact that these branches may refer to successor block which have not yet been visited by our transformation pass and therefore might be manipulated. Consequently, we created an intermediate dialect containing \emph{br} and \emph{cond\_br} operations that refer to blocks by their relative index, and a mapping between block indicies and the blocks themselves in the Flang generated IR is built. A separate rewrite is then run after the main transformation and leverages this mapping to replace these temporary operations with their \emph{cf.br} and \emph{cf.cond\_br} counterparts which point to their target specific blocks in the transformed IR.

\subsection{Handling of memory}

Flang represents variables via the \emph{hlfir.declare} operation, and this accepts as an input the memory which holds the value. This was observed in Listing \ref{lst:if_example_fir} where the function's integer argument of type (\emph{!fir.ref<i32>}) was provided to \emph{hlfir.declare}, and for local variables this memory is created via the \emph{fir.alloca} and \emph{fir.allocmem} operations which allocate on the stack and the heap respectively. Flang's generated IR is illustrated in Listing \ref{lst:var_example_fir}, and our transformation handles variables using the \emph{memref} dialect whose purpose is to allocate and manage memory, as well as providing the \emph{memref} type which represents the memory. The IR generated by Flang in Listing \ref{lst:var_example_fir} allocates an integer variable of type \emph{i32}, stores a value into it and then reads this back out. The \emph{fir.alloca} operation at line one reserves memory on the stack which is then packaged by the \emph{hlfir.declare} operation at line two. Lines four and five store the integer constant \emph{23} into the variable using the \emph{hlfir.assign} operation, and \emph{fir.load} at line seven loads the stored value.

\begin{lstlisting}[language=mlir, frame=lines, label=lst:var_example_fir, numbers=left, caption=HLFIR and FIR based IR generated by Flang which defines an integer scalar variable and iteracts with it]
%1 = fir.alloca() <{"bindc_name" = "i", "in_type" = i32}> : () -> !fir.ref<i32>
%2, %3 = hlfir.declare(%1) : (!fir.ref<i32>) -> (!fir.ref<i32>, !fir.ref<i32>)

%4 = arith.constant 23 : i32
hlfir.assign(%4, %2) : (i32, !fir.ref<i32>) -> ()
    
%5 = fir.load(%2) : (!fir.ref<i32>) -> i32
\end{lstlisting}

Listing \ref{lst:var_example_mlir} illustrates the IR generated by our mapping which transforms the IR in Listing \ref{lst:var_example_fir} into standard MLIR dialects. At line one it can be seen that we are using the \emph{memref.alloca} operation which, without any arguments, allocates a \emph{memref} of size one element. Lines three and four store the value 23 into this memref, using the \emph{memref.store} operation, and the value held in the memref is loaded at line six using the \emph{memref.load} operation. The empty square braces to the \emph{memref.store} and \emph{memref.load} operations denote that no array access indicies are provided as these are not required for single element memrefs.

\begin{lstlisting}[language=mlir, frame=lines, label=lst:var_example_mlir, numbers=left, caption=Standard MLIR dialect IR generated by our mapping lowering HLFIR and FIR based IR for allocating a scalar variable and interacting with it]
%1 = memref.alloca() : () -> memref<i32>

%2 = arith.constant 23 : i32
memref.store %2, %1[] : memref<i32>
    
%3 = memref.load %1[] : memref<i32>
\end{lstlisting}

Allocating arrays on the stack is similar to the above, where in Flang's generated IR the \emph{hlfir.declare} operation accepts an optional operand of type \emph{fir.shape} which itself has been created by the \emph{fir.shape} operation to define the size of the array. These are translated by our approach to the \emph{memref.alloca} operation, however we found that stack allocated memory was not being properly released. The MLIR documentation states that memref allocated stack memory is automatically released when control transfers back from the region of its closest surrounding operation with an \emph{AutomaticAllocationScope} trait, and that this trait applies to the \emph{func.func} operation. However, this seemed not to be the case and-so our transformation had to leverage the \emph{memref.alloca\_scope} operation which explicitly defines the scope of a stack frame. A restriction imposed by \emph{memref.alloca\_scope} is that its region can contain at most one block, which is incompatible with how our IR is laid out as described in Section \ref{sec:ctrl} as they use multiple blocks and jumps between them to support Fortran constructs such as \emph{gotos}. Consequently, we apply \emph{memref.alloca\_scope} around all function calls in our translated IR.

Allocatable arrays are represented by memrefs with dynamic size, for example the type \emph{memref<?x?xf64>} represents a dynamically sized two dimensional array of type double precision floating point. The \emph{memref.alloc} and \emph{memref.dealloc} operations are then used to allocate and deallocate these respectively on the heap. However, because allocatable arrays can be reallocated dynamically during program execution we must be able to refer to a pointer a memref, as allocation creates an entirely new memref. Therefore allocatable arrays are defined as memrefs that contain memrefs, where \emph{memref<memref<?x?xf64>>} is allocated on the stack for a two dimension allocatable array declaration. This creates an outer memref which contains, or points to, the inner memref which is the actual array.

\begin{lstlisting}[language=fortran, frame=lines, label=lst:array_example_ftn, numbers=left, caption=Example Fortran code which allocates an array and assigns a value to the second element]
integer, dimension(:), allocatable :: data
allocate(data(10))
data(2)=100
\end{lstlisting}

This is illustrated by the Fortran snippet in Listing \ref{lst:array_example_ftn}, where an array of size 10 is allocated and the value 100 assigned to the second element of this array. Flang then generates HLFIR and FIR based IR, which has been omitted for brevity, and this is then transformed by our approach into the standard MLIR dialects which is illustrated in Listing \ref{lst:array_example_mlir}. The \emph{memref.alloca} operation at line one of Listing \ref{lst:array_example_mlir} allocates the outer memref on the stack, with line two creating a constant of size 10 and this is provided as an argument to the \emph{memref.alloc} operation at line three which allocates the array itself of size 10 on the heap. This heap allocated memref, the array itself, is then stored at line four into the outer memref that was allocated on the stack at line one, which now contains, or points to, the memory allocated for the array.

\begin{lstlisting}[language=mlir, frame=lines, label=lst:array_example_mlir, numbers=left, caption=IR using MLIR's memref to represent an allocatable array]
%0 = "memref.alloca"() : () -> memref<memref<?xi32>>
%1 = arith.constant 10 : index
%2 = memref.alloc(%1) : memref<?xi32>
memref.store %2, %0[] : memref<memref<?xi32>>

%3 = arith.constant 100 : i32
%4 = arith.constant 2 : index
%5 = arith.constant 1 : index
%6 = arith.subi %4, %5 : index
%7 = memref.load %0[] : memref<memref<?xi32>>
memref.store %3, %7[%6] : memref<?xi32>
\end{lstlisting}

Lines six to eleven of Listing \ref{lst:array_example_mlir} handle the definition and storage of the integer value 100 into the array at index 2. In Fortran array indices start, by default, from 1 whereas memrefs follow C-style semantics and start at location zero. Therefore a subtraction of the index from its starting index is undertaken at line nine. The value being subtracted in this example is the default starting location of 1 because this has not been explicitly provided by the programmer. Our transformation tracks the starting location of each dimension of an array, either the default or programmer defined, and if this is known statically then it is represented by an \emph{arith.constant} as in Listing \ref{lst:array_example_mlir} or alternatively if it is dynamic then the SSA value is instead used in the subtraction. 

It can be seen in Listing \ref{lst:array_example_mlir} that line ten loads the allocated array memref which is contained by the outer memref, \emph{\%0}, and the value held in \emph{\%3} is then stored at line eleven to location \emph{\%6} of the array. Whenever an allocatable array is read from or stored to, first a \emph{memref.load} is issued, as per line ten, to extract the array's memref from its outer container. However, having to undertake this dereferencing on each array access potentially adds overhead. Consequently, we developed an optimisation pass which operates on the IR once it has been generated by our transformation, which identifies when the outer memref is updated. Specifically focused on loops, if the outer memref has not been updated then loads are reduced to a single operation and lifted out out of the loop, proceeding upwards through nests of loops as far as possible. Consequently in loops, which might involve a large number of array accesses, then stores and loads to the array can refer directly to the array memref rather than having to first load this from the outer memref container if the array is not reallocated. 

Our approach to handling allocatable arrays also conveniently represents Fortran pointers, and operations upon them. Furthermore, Fortran intrinsics such as \emph{movealloc} simply involve moving the array's memref between two outer memref containers. As memrefs contain their size, Fortran intrinsics to determine the size of arrays correspond to directly to the \emph{memref.dim} operation. Memrefs are also convenient when slicing arrays in Fortran because, for example, when passing a slice of an array to a procedure we can construct memref subviews and pass these. Subviews contain the same underlying memory as the original memref but allow for different offsets, sizes and strides, and-so this avoids copying sliced data into a new contiguous region when passing to a procedure.

Global arrays are represented by the \emph{memref.global} operation, which then uses \emph{memref.get\_global} in a function to load the memref from this global region. Global scalars are stored using \emph{llvm.mlir.global} with the body of this operation containing the \emph{arith.constant} operation that defines the value. These scalars are loaded using the \emph{llvm.mlir.addressof} operation to retrieve the LLVM pointer to the global, and is then dereferenced via \emph{llvm.load}.

\subsection{Handling of other constructs}

There are a variety of other Fortran constructs represented in HLFIR and FIR that must be converted to the standard MLIR dialects. Linear algebra intrinsics, such as \emph{dotproduct}, are one example and these are represented by explicit operations in HLFIR. Our translation pass lowers these to corresponding operations in the \emph{linalg} dialect, and Listing \ref{lst:sum_mlir} illustrates the MLIR IR generated to implement the Fortran \emph{sum} intrinsic which in this case is operating upon a one dimensional array. Line one allocates the output memref, here holding a single value which will be the result of the sum, that is then initialised to zero at lines two and three. Line four executes the \emph{linalg.reduce} operation on \emph{\%0} as an input (which is a memref of 10 elements of type \emph{i32}), with the output being the memref \emph{\%1} defined at line one. The \emph{dimensions} property defines which dimensions of the memref this reduction should operate over, and in this way we can encode the mask if this has been provided to the Fortran \emph{sum} intrinsic. Lines five to eight operate over each element of the input array and the current value held in the output, and add these together and yield this value at line seven which is then used in the next iteration. Line nine loads the resulting summed value in the output memref. In this manner, a range of other Fortran array intrinsics such as \emph{maxval} and \emph{product} are also be implemented.

\begin{lstlisting}[language=mlir, frame=lines, label=lst:sum_mlir, numbers=left, caption=IR using the linalg.reduce operation to implement the Fortran sum intrinsic]
%1 = "memref.alloca"() <{"operandSegmentSizes" = array<i32: 0, 0>}> : () -> memref<i32>
%2 = arith.constant 0 : i32
memref.store %2, %1[] : memref<i32>
linalg.reduce ins(%0:memref<10xi32>) outs(%1:memref<i32>) dimensions = [0] 
(%3 : i32, %4 : i32){
    %5 = arith.addi %3, %4 : i32
    linalg.yield %5 : i32
}
%6 = memref.load %1[] : memref<i32>
\end{lstlisting}

Flang represents derived types using the \emph{fir.type} attribute (type), which itself carries around the names of each member and their type. When referencing an element of this derived type Flang generates the \emph{hlfir.designate} operation which contains an optional a property called \emph{component} that refers to the member being accessed. Whilst this would map naturally to tuples in MLIR, MLIR lacks the constructs to manipulate these sufficiently. Consequently, for variables that are of a derived type we generate separate memrefs for each member and our transformation keeps track of which of these corresponds to which member of that derived type. 

\section{Performance}
\label{sec:performance}
Table \ref{tab:performance_comparison} reports performance of our approach (built using the latest release of LLVM), against the latest release of Flang, version 20.0.0, the Cray and GNU Gfortran compilers. It can be seen that our approach generally compares favourably against Flang and the performance delivered is mixed when compared to Gfortran. As per the results in Section \ref{sec:flang-perf-comparison}, the Cray compiler delivers very impressive performance against the others, however, crucially our approach is closing the gap using an LLVM/MLIR based solution, and this demonstrates the benefit of leveraging the MLIR ecosystem.

\begin{table}[htb]
    \centering
    \caption{Performance comparison between our approach, Flang v20, the Cray and GNU compilers.}
    \label{tab:performance_comparison}
    \begin{tabular}{|c|cccc|}
    \hline  
    & \multicolumn{4}{c|}{\textbf{Benchmark Runtime (s)}}\\
      \textbf{Benchmark} & \textbf{Our approach} & \textbf{Flang v20} &  \textbf{Cray} &\textbf{GNU}\\
      \hline
    ac & 10.23 & 11.89 & 8.67 & 31.43 \\
    linpk & 5.43 & 6.24 & 5.79 & 4.81 \\
    nf & 10.69& 10.29 & 7.72 & 7.43 \\
    test\_fpu & 72.41 & 110.80 & 32.56 & 76.99 \\
    tfft & 52.33 & 48.90 & 61.65 & 115.86 \\
    jacobi & 249.08 & 277.67 & 109.89 & 232.62 \\
    pw-advection & 86.47 & 205.33 & 47.28 & 192.05 \\
    tra-adv & 124.72 & 141.95 & 79.38 & 116.71 \\
    \hline
    \end{tabular}
\end{table}

The \emph{jacobi}, \emph{pw-advection} and \emph{tra-adv} benchmarks especially required tuning and motivated the development of additional optimisation passes. For example, the \emph{pw-advection} benchmark was initially approximately four times slower than the final performance reported in Table \ref{tab:performance_comparison}, and there were two important optimisations that made a crucial difference across all these benchmarks. Firstly, the use of dynamically sized memrefs, for example \emph{memref<?x?xf64>}, severely impacted the effectiveness of many of the MLIR optimisation passes and this caused poor performance. We therefore developed an optimisation pass which operates on the IR that is generated by our transformation and identifies whether allocatable arrays are allocated by statically known bounds and then not reallocated throughout a procedure. If so, then this optimisation pass will replace the dynamically sized memref, \emph{memref<?x?xf64>}, with it's static counterpart, for instance \emph{memref<128x128xf64>}. It also must rewrite the \emph{memref.alloc} operation to remove the dynamic bound operands, instead encoding the static bounds in the result type.

The other important optimisation was to vectorise loops. We initially experimented with the \emph{scf-for-loop-specialization} optimisation pass, but this was ineffective and did not vectorise the code. Instead, we developed a transformation pass which operates upon the standard IR resulting from our transformation, to convert \emph{scf.for} loops into their affine counterparts where possible. Using the \emph{affine} dialect, this involved not only converting the loop constructs into \emph{affine.for}, but furthermore rewriting memref loads and stores to the corresponding operations in the affine dialect, using the loop variable directly as indexes with optional constant offsets. There are a rich set of loop optimisation passes provided by MLIR for affine loops, including vectorisation, tiling and unrolling. 

\begin{figure}[htb]
\centering
 \includegraphics[width=\columnwidth]{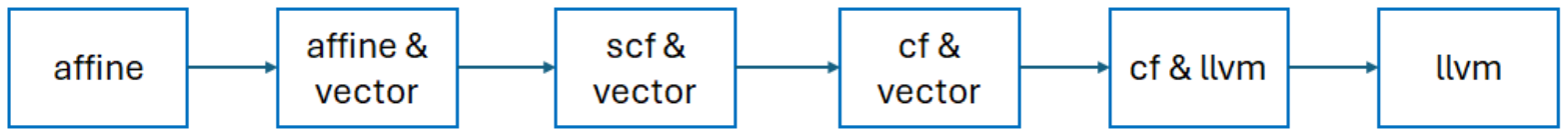}
\caption{Illustration of applying vectorisation to the affine dialect and then lowering to the LLVM dialect}	
\label{fig:vector-flow}
\end{figure}

Figure \ref{fig:vector-flow} illustrates this flow, where we used the pass \emph{affine-super-vectorize\{virtual-vector-size=4\}} to apply vectorisation, and the size we selected is four because the AMD Rome CPU in ARCHER2 is AVX2 providing a 256 bit width. The affine dialect is then lowered to the \emph{scf} dialect, which is lowered to the \emph{cf} dialect and at this point we execute the pass \emph{convert-vector-to-llvm\{enable-x86vector\}} to lower the \emph{vector} dialect to the \emph{llvm} dialect targeting vector instructions for x86. Lastly, the \emph{cf} dialect is lowered to the \emph{llvm} dialect.

\subsection{Linear Algebra operations}
Fortran provides, as intrinsics, numerous linear algebra operations to the programmer. However, whilst there is a linear algebra dialect in MLIR called \emph{linalg}, Flang does not use this and instead provides support in the runtime library which these intrinsics call into. By contrast, the \emph{linalg} dialect provides a wide range of operations which are lowered by MLIR transformation passes into a choice of other dialects including \emph{affine} and \emph{scf}. Our transformation approach from Flang's HLFIR and FIR into the standard MLIR dialects leverages the \emph{linalg} dialect for these intrinsics, and an important question is which approach delivers better performance.

Table \ref{tab:linear_alg_comparison} reports a performance comparison between Flang and our approach for four common intrinsic operations which map directly to operations in the \emph{linalg} dialect when using our approach, or alternatively call functions in the Fortran runtime when using Flang. The first of these is transposing a two dimension integer array of size 32768 by 32768, the second undertakes matrix multiplication on double precision floating point matrices of size 4096 by 4096, the third performs dot product on double precision floating point vectors of size 134 million elements, and the fourth calculates the sum of the elements held in a two dimensional double precision array of size 32768 by 32768. Comparing the serial performance of our approach against Flang in Table \ref{tab:linear_alg_comparison}, it can be seen that leveraging the \emph{linalg} dialect always delivers better performance compared to the runtime library approach of Flang. 

\begin{table}[htb]
    \centering
    \caption{Performance of Fortran intrinsics in our approach and Flang, using the \emph{linalg} dialect in our approach compared to runtime library functions used by Flang.}
    \label{tab:linear_alg_comparison}
    \begin{tabular}{|c|cc|c|}
    \hline     
      \textbf{Operation} & \multicolumn{2}{c|}{\makecell{\textbf{Our approach} \\ \textbf{runtime (s)}}} & \makecell{\textbf{Flang v20} \\ \textbf{runtime (s)}}\\
      & \textbf{serial} & \textbf{threaded} & \textbf{serial}\\
      \hline
    transpose & 214.48 & 40.75 & 272.38 \\
    matmul & 43.12 & 11.85 & 45.71 \\
    dotproduct & 0.81 & - & 2.70 \\
    sum & 1.63 & - & 1.65\\
    \hline
    \end{tabular}
\end{table}

We lower the \emph{linalg} dialect to \emph{affine}, and both the \emph{matmul} and \emph{dotproduct} benchmarks benefited the greatest from affine optimisation passes. For instance, the runtime of \emph{matmul} was initially around five times slower and to deliver the performance reported in Table  \ref{tab:linear_alg_comparison} loop tiling was undertaken using the \emph{affine-loop-tile} pass. Likewise, loop unrolling and vectorisation resulted in around a two times performance improvement for the \emph{dotproduct} benchmark. The \emph{threaded} column of our approach in Table \ref{tab:linear_alg_comparison} demonstrates another benefit of using the standard \emph{linalg} dialect. We developed a very simple transformation pass that converts appropriate \emph{scf.for} loops to their \emph{scf.parallel} loop counterparts, enabling us to then target OpenMP via MLIR's \emph{convert-scf-to-openmp} transformation which lowers parallel loops to the OpenMP dialect. Our current transformation is very simple and does not yet fully support reductions, hence we have not yet run this on the \emph{dotproduct} or \emph{sum} benchmarks in this section. However, over 64 cores this delivers performance improvements for both the \emph{transpose} and \emph{matmul} operations and demonstrates the benefit and flexibility of these composable dialects and transformations. It should be noted that there is a transformation directly from the \emph{linalg} dialect to \emph{scf.parallel} loops, however this bypasses the \emph{affine} dialect and as discussed in this section that offers important performance optimisations.

\subsection{OpenMP parallelism}

Whilst Flang leverages the OpenMP, \emph{omp}, MLIR dialect internally it provides its own lowering to call into the OpenMP runtime library. By contrast, there is a \emph{convert-openmp-to-llvm} MLIR pass which lowers the \emph{omp} dialect to the \emph{llvm} dialect. Our translation pass conserves IR using the \emph{omp} dialect generated by Flang, and then leverages the MLIR transformation to lower this. A question is therefore whether there is any performance difference between these two approaches.

We added OpenMP loop pragmas to the \emph{jacobi} and \emph{pw-advection} benchmarks and compared the performance delivered by our approach and Flang. Table \ref{tab:openmp} reports the speed up n a specific number of cores compared to the serial code, for these two benchmarks. It can be seen that these two approaches are comparable for the \emph{pw-advection} benchmark, but for the \emph{jacobi} benchmark leveraging MLIR delivers greater performance when running on larger numbers of cores. Given that these both ultimately call into the OpenMP runtime library, the magnitude of this difference was unexpected and to understand this we explored the LLVM-IR that is generated by both approaches. Whilst both approaches call the same OpenMP runtime functions there are some significant differences, for instance the MLIR generated approach issues \emph{llvm.stacksave} on entry to and \emph{llvm.stackrestore} on exit from the the worksharing loop, whereas the Flang generated code does not. However the greatest difference is in the loop body itself, where for each iteration the MLIR generated code comprises 29 instructions, but this is 80 instructions for that generated by Flang. Consequently whilst we had, somewhat naively, assumed that the generated code would be very similar as the OpenMP runtime library provides much of the threading support, in-fact there are significant differences between MLIR and Flang around how the code that is provided to these library calls is structured.

\begin{table}[htb]
    \centering
    \caption{Speed up compared to serial execution when threading using OpenMP for our approach and Flang}
    \label{tab:openmp}
    \begin{tabular}{|c|cc|cc|}
    \hline     
      \textbf{Number} & \multicolumn{2}{c|}{\textbf{Our approach speedup}} & \multicolumn{2}{c|}{\textbf{Flang v20 speedup}} \\
     \textbf{of cores} & \textbf{jacobi} & \textbf{pw-advection} & \textbf{jacobi} & \textbf{pw-advection} \\
      \hline
    2 & 1.95 & 1.81  & 1.76 & 1.82 \\
    4 & 4.01 & 3.34 & 3.42 & 3.28 \\
    8 & 5.77 & 5.52 & 6.47 & 5.37 \\
    16 & 13.14 & 8.04 & 11.43 & 7.75\\
    32 & 26.14 & 9.77 & 13.96 & 9.75\\
    64 & 72.62 & 10.80 & 18.39 & 10.90\\
    \hline
    \end{tabular}
\end{table}

\subsection{GPU acceleration}
Flang supports acceleration of codes on GPUs using both OpenACC and OpenMP. MLIR provides these as dialects, as well as a general \emph{gpu} dialect in addition to vendor specific ones. We therefore added OpenACC pragmas into the \emph{pw-advection} benchmark to experiment with generating GPU code from our approach. Somewhat surprisingly, there is no MLIR pass to lower from the OpenACC, \emph{acc}, dialect. Consequently we developed our own transformation operating on the IR generated by our approach and changes all \emph{scf.for} loops that are within an \emph{acc.kernels} operation into \emph{scf.parallel} loops. We then leverage the existing MLIR transformation \emph{convert-parallel-loops-to-gpu} to lower these parallel loops to the \emph{gpu} dialect and subsequent passes lower this further and generate the GPU binary which is embedded within the IR. In this initial version we leverage CUDA managed memory, and consequently our transformation pass also converts the \emph{acc.create} operation into \emph{gpu.host\_register} and also inserts \emph{gpu.host\_unregister} at the end replacing \emph{acc.delete} and \emph{acc.copyout}.

Unfortunately the version of CUDA on our GPU machine limited the version of Flang that we could build, and we were restricted to version 18 of Flang (part of the v17.0.6 release of LLVM). When using Flang to build OpenACC code the compiler threw an internal error reporting a missing \emph{LLVMTranslationDialectInterface} and then segmentation faulted. This was during the translation of HLFIR and FIR IR into LLVM-IR, and-so did not effect our approach. 

\begin{table}[htb]
    \centering
    \caption{Runtime on a Nvidia V100 GPU of the \emph{pw-advection} benchmark for different grid sizes using OpenACC compiled with our approach and Nvidia's Fortran compiler}
    \label{tab:gpu}
    \begin{tabular}{|c|c|c|}
    \hline     
      & \multicolumn{2}{c|}{\textbf{Benchmark Runtime (s)}}\\
     \textbf{Number of grid cells} & \textbf{Our approach} & \textbf{nvfortran} \\
      \hline
    134 million & 4.72 & 3.88 \\
    268 million & 6.33 & 5.94 \\
    536 million & 11.65 & 10.84  \\
    1.1 billion & 22.78 & 21.80 \\
    \hline
    \end{tabular}
\end{table}

Consequently, we compared against Fortran OpenACC code built using the Nvidia Fortran compiler (version 22.11) and compared our approach against that. The results of this experiment on a V100 GPU in the Cirrus supercomputer, for the \emph{pw-advection} benchmark, are reported in Table \ref{tab:gpu}. Whilst it can be seen that the Nvidia compiler outperforms our approach, arguably they are fairly close which is impressive given that we are mostly leveraging existing passes, no specific optimisation, and adding GPU acceleration to our transformation was trivial based upon existing MLIR support. Indeed, in our opinion this further demonstrates the benefit of leveraging standard MLIR because not only does this already provide many of the required building blocks, but furthermore future optimisation passes can manipulate the IR at any point in the compilation pipeline to deliver improved tuning and performance. 

\section{Conclusions and further work}
\label{sec:conclusions}
In this paper we have explored an alternative compilation approach for Flang which was driven by the motivation discovered in Section \ref {sec:flang-perf-comparison} that there is still someway to go in matching the performance of the Cray and GNU Fortran compilers. One of the potential limitations of Flang is that it undertakes a bespoke lowering from its HLFIR and FIR dialects, which are mixed with a handful of standard dialects, directly to LLVM-IR. By contrast, we have developed a transformation that generates standard MLIR dialects from HLFIR and FIR, enabling the compilation flow to then benefit from the wide range of MLIR dialects and transformations that are available.

It was our hypothesis that integration with standard MLIR would provide improved performance by leveraging the existing work of the MLIR community. We described our approach and the mixing of dialects to represent different Fortran constructs, as well as demonstrating that MLIR provides a wide range of dialects and passes to choose from. We demonstrated that, in the main, our approach outperforms Flang and in some cases also out performs GNU's Gfortran although falls someway short of the Cray compiler. Furthermore, we also demonstrated that whilst the two approaches to providing linear algebra operations, libraries in Flang compared to the \emph{linalg} dialect in MLIR, deliver broadly similar performance, the MLIR approach is slightly faster and indeed provides increased flexibility. Lastly, when exploring threading performance with OpenMP we found that whilst the speed up delivered between Flang and our approach is generally comparable, there were significant benefits in using the standard MLIR flow for the Jacobi benchmark at larger core counts.

The work described in this paper is a research prototype and has been developed to explore the overarching question of whether this is a viable and worthwhile approach. Based on these results it is now our intention to further develop and extend this, for instance improving support for string handling, supporting a wider range of applications, and further performance tuning. We also plan on further exploring GPU performance not only with OpenACC but also using OpenMP offload. Clearly there are opportunities for optimisation in our GPU offload implementation, and there are a large number of operations in MLIR's \emph{gpu} dialect that we can leverage and then run across both Nvidia and AMD GPUs. 

We therefore conclude that integrating Flang with standard MLIR is advantageous. The work being undertaken on the Flang compiler is impressive, however based upon the results in this paper it is our belief that the community should seriously explore this alternative compilation pipeline and consider integrating it as an optional flow into Flang.

\section*{Acknowledgments}
This research was funded by the ExCALIBUR EPSRC xDSL project grant number EP/W007940/1. This research was supported by an RSE personal research fellowship. This work used the ARCHER2 UK National Supercomputing Service (https://www.archer2.ac.uk). For the purpose of open access, the author has applied a Creative Commons Attribution (CC BY) licence to any Author Accepted Manuscript version arising from this submission.

\bibliographystyle{IEEEtran}
\bibliography{references.bib}

\end{document}